\documentclass[a4paper]{article}
\usepackage[margin=25mm]{geometry}
\usepackage{url}
\usepackage{import}
\usepackage{amsfonts,amssymb,amsmath,amsthm,dsfont}
\usepackage{xcolor}
\usepackage{comment}
\usepackage[linesnumbered,algoruled]{algorithm2e}
\usepackage{bm}
\usepackage[normalem]{ulem}
\usepackage{graphicx}
\graphicspath{{figures/}}
\usepackage{adjustbox}
\usepackage{mathtools}

\providecommand{\keywords}[1]
{
  \small
  \textbf{\textit{Keywords---}} #1
}

\newcommand{\prox}{\text{prox}}
\newcommand{\APL}{\textsf{APL}}
\newcommand{\tildeAPL}{\widetilde{\textsf{APL}}}
\newcommand{\argmin}[1]{\text{arg}\underset{#1}{\text{min}}}

\begin{document}

\title{Learning the Proximity Operator in Unfolded ADMM \\for Phase Retrieval\thanks{This work is supported by the European Research Council (ERC FACTORY-CoG-6681839) and by ANITI under grant agreement ANR-19-PI3A-0004.}}
\date{}
\author{Pierre-Hugo~Vial\thanks{IRIT, Université de Toulouse, CNRS, Toulouse, France (e-mail: firstname.lastname@irit.fr).} ,
        Paul~Magron\thanks{Université de Lorraine, CNRS, Inria, LORIA, F-54000 Nancy, France (e-mail: firstname.lastname@inria.fr).} ,
        Thomas~Oberlin\thanks{ISAE-SUPAERO, Université de Toulouse, France (e-mail: firstname.lastname@isae-supaero.fr).} ,
        Cédric~Févotte\footnotemark[2]
}

\maketitle
\begin{abstract}
This paper considers the phase retrieval (PR) problem, which aims to reconstruct a signal from phaseless measurements such as magnitude or power spectrograms. PR is generally handled as a minimization problem involving a quadratic loss. Recent works have considered alternative discrepancy measures, such as the Bregman divergences, but it is still challenging to tailor the optimal loss for a given setting. In this paper we propose a novel strategy to automatically learn the optimal metric for PR. We unfold a recently introduced ADMM algorithm into a neural network, and we emphasize that the information about the loss used to formulate the PR problem is conveyed by the proximity operator involved in the ADMM updates. Therefore, we replace this proximity operator with trainable activation functions: learning these in a supervised setting is then equivalent to learning an optimal metric for PR. Experiments conducted with speech signals show that our approach outperforms the baseline ADMM, using a light and interpretable neural architecture.
\end{abstract}
\section{Introduction}
\label{sec:intro}

Phase retrieval (PR) consists in reconstructing data from phaseless nonnegative measurements. This problem finds practical applications in various areas such as optical imaging~\cite{walther63, harrison93} and audio signal processing~\cite{Gerkmann2015, Mowlaee2016}, which is the domain of interest of this paper. Traditionally, PR is formulated as the following optimization problem:
\begin{equation}
    \label{eq:pr}
    \underset{\mathbf x \in \mathbb R^L}{\text{min}} \| |\mathbf A \mathbf x|^d - \mathbf r\|^2,
\end{equation}
where $\mathbf A \in \mathbb{C}^{K \times L}$ is the measurement operator, $\mathbf r \in \mathbb{R}_+^K$ are the phaseless measurements, and $||.||$ denotes the Euclidean norm. In audio signal processing, $\mathbf A$ is often the short-time Fourier transform (STFT), and the measurements are either magnitude ($d=1$) or power ($d=2$) spectrograms. In the seminal work \cite{GLA}, the authors consider STFT magnitude measurements and propose an iterative procedure, known as the Griffin-Lim algorithm (GLA) which is proved to converge to a critical point of \eqref{eq:pr}. Other optimization algorithms were proposed to tackle Problem \eqref{eq:pr}, such as majorization-minimization~\cite{PRIME}, gradient descent~\cite{WF} or alternating direction method of multipliers (ADMM)~\cite{liang17}. PR has also been addressed by replacing the quadratic loss in~\eqref{eq:pr} with alternative discrepancy measures such as Bregman divergences~\cite{vial2021phase}, which have been shown to lead to superior results in many audio applications~\cite{fevotte09, LeRoux2011}. However, prescribing a loss that is optimal for all signal processing problems and classes of audio signals remains challenging~\cite{vial2021phase}. 

On the other hand, recent PR approaches have leveraged deep neural networks (DNNs)~\cite{Takamichi2018, Thieling2021, Thien2021, arik2018fast}. Despite their successful performances in a large number of tasks, the enthusiasm for DNNs can be tempered by a general lack of explainability due to their black box structure, and by their limited ability to generalize to unseen data or experimental conditions. Deep unfolding (or unrolling)~\cite{gregor2010learning, hershey2014deep} is a promising attempt to alleviate these limitations with model-based architectures derived from iterative algorithms. This strategy consists in considering each iteration of an optimization algorithm as a (trainable) layer of a DNN. It has proved very promising when applied to many algorithms in signal processing \cite{hershey2014deep,bertocchi2020deep, monga2021algorithm}, including audio PR~\cite{Masuyama2019deepGLA, Wang2018a, Wichern2018}. However, these approaches treat the linear layers as trainable parameters while activation functions remain fixed, although recent studies show that trainable activation functions can upgrade performance~\cite{qian2018adaptive, apicella2019simple, scardapane2019kafnets}. Besides, they do not address the problem of learning the metric involved in the PR formulation problem.

In this paper, we propose to unfold the ADMM algorithm for PR proposed in~\cite{vial2021phase}. Our method builds upon observing that the choice of the discrepancy measure only affects the computation of a proximity operator in the ADMM updates. Therefore, we can recast the problem of metric learning as a problem of proximity operator learning in the unfolded ADMM. To that end, we replace this proximity operator with a trainable activation function. We show that the proposed parametrization of the network is connected to the metric involved in the original optimization problem, which yields an interpretable architecture. Experiments performed on speech signals demonstrate the efficiency of our method, which outperforms a baseline ADMM~\cite{liang17} with a number of iterations equal to the number of layers in the unfoldeed ADMM.

The rest of this paper is structured as follows. Section~\ref{sec:related_work} presents the related work. Then, the proposed method is introduced in Section~\ref{sec:method} and tested experimentally in Section~\ref{sec:exp}. Finally, Section~\ref{sec:conc} draws some concluding remarks.

\vspace{0.5em}

\noindent \textbf{Mathematical notation}:

\begin{itemize}
    \item $\mathbf{A}$ (capital, bold font): matrix.
    \item $\mathbf{x}$ (lower case, bold font): vector.
    \item $z$ (regular): scalar.
    \item $|.|$, $\angle(.)$: magnitude and complex angle, respectively.
    \item $\mathbf{x}^\mathsf{H}$: Hermitian transpose.
    \item $\odot$, $(.)^d$, fraction bar: element-wise matrix or vector multiplication, power, and division, respectively.
\end{itemize}

\section{Related work}
\label{sec:related_work}
\subsection{PR with Bregman divergences}

In \cite{vial2021phase}, we proposed to reformulate the PR problem by substituting the quadratic loss in~\eqref{eq:pr} with a Bregman divergence. Bregman divergences encompass beta-divergences~\cite{Hennequin2011}, which include the Kullback-Leibler and Itakura-Saito divergences, as well as the quadratic loss. They are defined as follows:
\begin{equation}
\mathcal D_\psi (\mathbf p\, \bm | \, \mathbf q)= \sum_{k=1}^K \left[ \psi(p_k) - \psi(q_k) - \psi'(q_k)(p_k - q_k) \right],
\label{eq:breg}
\end{equation}
where $\psi$ is a strictly-convex, continuously-differentiable generating function (with derivative $\psi'$). Since Bregman divergences are not symmetric in general, we considered the two following problems, respectively termed ``right" and ``left":
\begin{equation}
    \underset{\mathbf x \in \mathbb R^L}{\text{min}}\, \mathcal D_\psi(\mathbf r \,\bm |\, |\mathbf A \mathbf x|^d) \text{ and } \underset{\mathbf x \in \mathbb R^L}{\text{min}}\, \mathcal D_\psi(|\mathbf A \mathbf x|^d \,\bm |\, \mathbf r).
    \label{eq:bregpr_right}
\end{equation}
Two algorithms were derived for solving these problems, based on gradient descent and ADMM~\cite{vial2021phase}.

\subsection{PR via ADMM}
\label{sec:PRviaADMM}

In this work we focus on the ``left" PR problem using ADMM, since it produced the best reconstruction results in~\cite{vial2021phase}. Drawing on~\cite{liang17}, this approach consists in reformulating~\eqref{eq:bregpr_right} into the following constrained optimization problem:
\begin{equation}
        \underset{\mathbf x \in \mathbb R^L, \mathbf u \in \mathbb R^K_+, \theta \in \left [0;2 \pi \right [^K}{\text{min}}\, \mathcal D_\psi(\mathbf u \,\bm|\, \mathbf r) \quad \mbox{s.t.} \quad (\mathbf A \mathbf x)^d = \mathbf u \odot e^{i\bm \theta},
        \label{eq:bregpr_admm}
\end{equation}
where $\mathbf u$ and $\bm \theta$ are auxiliary variables for the magnitude and phase of $\mathbf{A}\mathbf{x}$. From~\eqref{eq:bregpr_admm} we obtain the augmented Lagrangian:
\begin{equation}
    \mathcal L(\mathbf x, \mathbf u, \bm \theta, \bm \lambda) = \mathcal D_\psi(\mathbf u \,\bm|\, \mathbf r) + \mathfrak R \left( \bm \lambda^{\mathsf H} ((\mathbf A\mathbf x)^d-\mathbf u \odot e^{i\bm \theta}) \right) \nonumber +\frac{\rho}{2}\left\|(\mathbf A\mathbf x)^d-\mathbf u \odot e^{i\bm \theta}\right\|^2,
\end{equation}
where $\mathfrak R$ is the real part function, $\bm \lambda \in \mathbb{C}^K$ are the Lagrange multipliers and $\rho >0$ is the augmentation parameter. Alternate minimization of $\mathcal{L}$ leads to the following update rules \cite{vial2021phase}:
\begin{align}
    \label{eq:admm_h}
    \mathbf h_{t+1} &=(\mathbf A \mathbf x_t)^d + \frac{\bm \lambda_t}{\rho}
\\
  \mathbf u_{t+1} &= \text{prox}_{\rho^{-1}\mathcal D_\psi(\cdot \, \bm | \, \mathbf r)}(|\mathbf h_{t+1}|)
  \label{eq:ADMM_u_theta}
\\
  \bm \theta_{t+1}&=\angle \mathbf h_{t+1}
\\
    \label{eq:admm_x}
    \mathbf x_{t+1}&=\mathbf A^\mathsf H\big ( \mathbf u_{t+1}\odot e^{i \bm \theta_{t+1}} -\frac{\bm \lambda_t}{\rho} \big ) ^{1/d}
\\
    \label{eq:admm_lambda}
    \bm \lambda_{t+1} &= \bm \lambda_{t} + \rho (\mathbf A \mathbf x_{t+1} - \mathbf u_{t+1}\odot e^{i \bm \theta_{t+1}}),
\end{align}
where $\text{prox}_f$ denotes the proximity operator of a function $f$, defined as the mapping of a vector $\mathbf y$ to the solution of the minimization problem:
\begin{equation}
    \prox_{\rho^{-1}f}(\mathbf y) := \argmin{\mathbf z \in \mathbb R^K} f(\mathbf z) + \frac{\rho}{2}\|\mathbf z - \mathbf y\|^2.
\end{equation}
However, a closed-form expression of this operator is not available for every Bregman divergence. To alleviate this issue, we propose in this paper to replace this proximity operator with a learnable neural activation function, as detailed in the next section.

\section{Proposed method}
\label{sec:method}

\subsection{General architecture}

\begin{figure}[t]
    \centering
    \includegraphics[scale=0.4]{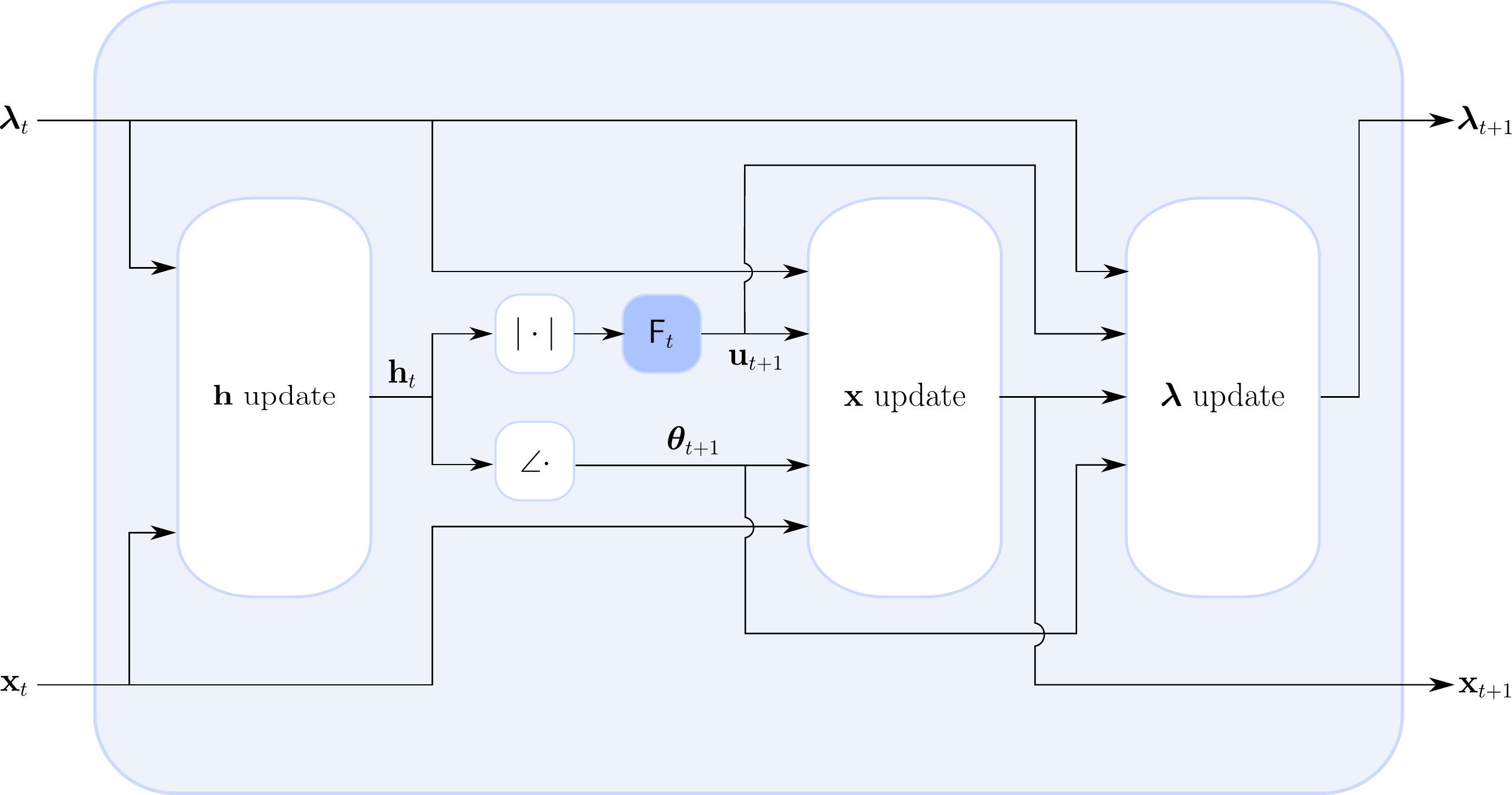}
    \caption{One layer of the proposed unfolded architecture.}
    \label{fig:archi}
\end{figure}

The ADMM updates detailed in~\ref{sec:PRviaADMM} consist in successive linear and nonlinear computations. As such, this algorithm can be viewed as a neural network $\bm{\mathsf U}$ via unfolding:
\begin{equation}
    (\mathbf x_T,\, \bm \lambda_T) = \bm {\mathsf U}(\mathbf x_0, \,\bm \lambda_0) = \mathsf U_1 \circ \dots \circ \mathsf U_T(\mathbf x_0, \,\bm \lambda_0),
\end{equation}
where $\mathsf U_t$ denotes the $t$-th layer of the network, mimicking the $t$-th iteration of the ADMM algorithm, as illustrated in Fig.~\ref{fig:archi}. 
The layer $\mathsf U_t$ can be decomposed into two linear parts denoted by $\mathsf L_t^{(1)}$ and $\mathsf L_t^{(2)}$, and a nonlinear part $\mathsf{NL}_t$ as follows:
\begin{align}
    \mathsf L_t^{(1)} &: (\mathbf x_{t-1},\,\bm \lambda_{t-1}) \mapsto \mathbf h_t \\ 
    \mathsf{NL}_t &: \mathbf h_t \mapsto (\mathbf u_t,\, \bm \theta_t) = \left (\mathsf F_t(|\mathbf h_t|,\, \mathbf r),\, \angle \mathbf h_t\right) \label{eq:nl} \\
    \mathsf L_t^{(2)} &: (\mathbf x_{t-1},\,\bm \lambda_{t-1}, \mathbf u_t,\, \bm \theta_t) \mapsto (\mathbf x_t,\, \bm \lambda_t),
\end{align}
with $\mathbf h_t, \mathbf x_t,\, \bm \lambda_t$ respectively defined as in \eqref{eq:admm_h}, \eqref{eq:admm_x} and \eqref{eq:admm_lambda}. $\mathsf F_t$ denotes a parameterized sublayer modeling the proximity operator of equation \eqref{eq:ADMM_u_theta}. Since the choice of the discrepancy measure $\mathcal{D}_{\psi}$ only affects the proximity operator~\eqref{eq:ADMM_u_theta} in the updates, we can recast the problem of metric learning as the problem of proximity operator learning. We propose to leverage a trainable activation function in order to model this layer and learn the proximity operator.

\subsection{Proposed parameterization}
\label{sec:method_parametrization}

To build the non-linear sublayers $\mathsf F_t$ that model $\prox_{\rho^{-1}\mathcal D_{\psi(\cdot\,|\,\mathbf r)}}$, we first reformulate this operator as follows. Let $\mathbf v\in \mathbb R^K$ and $f(\mathbf z) =  \sum_{k=1}^K [\psi(z_k) + v_k z_k]$. We have~\cite{combettes2005signal}: 
\begin{equation}
    \prox_{\rho^{-1}f} (\mathbf y)= \prox_{\rho^{-1}\tilde \psi}\left(\mathbf y-\rho^{-1}\mathbf v\right),
    \label{eq:prop_prox}
\end{equation}
where $\tilde \psi(\mathbf z) = \sum_k \psi(z_k)$. Setting $\mathbf v = - \psi'(\mathbf r)$ in~\eqref{eq:prop_prox}, with $\psi'$ applied entrywise, it is straightforward to see that:
\begin{equation}
    \prox_{\rho^{-1}\mathcal D_\psi(\cdot\,\bm|\,\mathbf r)}(\mathbf y) = \prox_{\rho^{-1}\tilde \psi}(\mathbf y + \rho^{-1} \psi'(\mathbf r)).
    \label{eq:prox_reformulation}
\end{equation}
This formulation of the proximity operator is more convenient than~\eqref{eq:ADMM_u_theta} since the measurements $\mathbf{r}$ no longer appear in the input function of the proximity operator, but instead in the argument of the latter (with $\mathbf{y}$). This leads to a more natural parametrization for unrolling.

Let us first derive the proximity operator \eqref{eq:prox_reformulation} in a simple scenario, namely the quadratic loss ($\tilde \psi = \frac 1 2 \|\cdot\|^2$, $\psi'(\mathbf r)=\mathbf r$). In this case we have~\cite{liang17}:
\begin{equation}
    \prox_{\rho^{-1}\frac{1}{2}\|\cdot-\mathbf r\|^2}(\mathbf y) = \frac{\mathbf y + \rho^{-1}\mathbf r}{1+\rho^{-1} }.
    \label{eq:prox_quad}
\end{equation}
As a result, a first simple approach for proximity operator learning would consist in treating $\rho$ as a learnable parameter.
However, early experiments have shown poor performance with this approach, which is due to the very low expressive power of such a model (only one scalar value).
More generally, one can consider a beta-divergence with shape parameter $\beta$, for which $\psi'(\mathbf r) = \frac{\mathbf r^{\beta-1}}{\beta -1}$~\cite{Hennequin2011}. However, the proximity operator of $\tilde \psi$ is not available for every beta-divergence.

To alleviate this issue, we model this unavailable proximity operator using Adaptive Piecewise Linear  (APL) activations~\cite{agostinelli2015learning}. They are defined by:
\begin{equation}
    \APL(\mathbf y) := \max(\mathbf y, 0)  + \sum_{c=1}^C w_c \max(-\mathbf y+b_c, 0),
    \label{eq:apl}
\end{equation}
where $w_c$ and $b_c$ are learnable parameters controlling the slopes and biases of the linear segments, and the $\max$ is applied entry-wise. Then, we propose the following parametrization of the nonlinear layer $\mathsf F_t$:
\begin{equation}
    \mathsf F_t(\mathbf y, \, \mathbf r)=\APL_t\left(\gamma^{(1)}_t\mathbf y + \gamma^{(2)}_t\frac{\mathbf r^{\beta_t-1}}{\beta_t-1}\right),
    \label{eq:prox_param}
\end{equation}
with learnable parameters $w_{c,t}, b_{c,t}$, $\gamma^{(1)}_t$, $\gamma^{(2)}_t$, and $\beta_t$. Even though \textit{ad hoc}, this parametrization is motivated by the following considerations:
\begin{itemize}
    \item APL can represent any continuous piecewise linear function over a subset of real numbers. As such, it generalizes the proximity operator obtained in the quadratic case~\cite{liang17}.
    \item The term in the form of $\frac{\mathbf r^{\beta-1}}{\beta-1}$ in~\eqref{eq:prox_param} is reminiscent of $\psi'$ for beta-divergences, as mentioned above.
    \item Introducing learnable weights $\gamma^{(1)}_t$ and $\gamma^{(2)}_t$ allows to increase the model capacity, as it was shown beneficial in our preliminary experiments.
\end{itemize}
Note that when $w_c=0,\, \gamma_t^{(1)}=\frac{1}{1+\rho^{-1}},\, \gamma_t^{(2)}=\frac{\rho^{-1}}{1+\rho^{-1}}$ and $\beta_t=2$, $\mathsf F_t$ is equal to the proximity operator for the quadratic loss~\eqref{eq:prox_quad}. Overall, our parametrization~\eqref{eq:prox_param} is a good trade-off between tractability, interpretability, and expressiveness. 

Two variants of the proposed architecture will be considered in our experiments. In the ``untied" variant, each layer uses different parameters, and the global set of parameters is $\left\{ \{  w_{c,t}, b_{c,t} \}_{c=1}^C , \gamma^{(1)}_t, \gamma^{(2)}_t, \beta_t\right\}_{t=1}^T$, while in the ``tied" variant, the parameters are shared among layers, i.e., constant with $t$.

In the end, after learning these parameters (see Section~\ref{sec:training}), the proposed method, termed unfolded ADMM (UADMM), estimates a signal $\mathbf{x}_T$ via $(\mathbf{x}_T, \bm \lambda_T) = \bm{\mathsf{U}}(\mathbf{x}_0, \bm \lambda_0)$, where $\mathbf{x}_0$ is some initial estimate.

\subsection{Discussion about interpretability}
\label{sec:method_interpretation}

Under mild assumptions and as detailed in the supplementary material,\footnote{In particular, the APL unit must be strictly increasing, something we ensure by imposing the weights to be negative, using $ w_{c,t}= - \tilde w_{c,t}^2$.} we can prove that there exists a close-form function $f_{\mathbf r,t}: \mathbb R^K \rightarrow \mathbb R \cup \{+\infty\}$ such that ${\mathsf F_t(\mathbf y, \mathbf r)=\prox_{f_{\mathbf r,t}}(\mathbf y)}$, 
(see~\eqref{eq:inv_prox} in the supplementary material). In the ``tied" variant, where $f_{\mathbf r,t}=f_{\mathbf r}$, reconstructing $f_{\mathbf r}$ from the learned parameters is analogous to identifying the metric $\mathcal D_\psi(\cdot\,|\, \mathbf r)$ involved in the PR optimization problem. With the relaxation proposed in the ``untied" case, this interpretation is more limited
as $f_{\mathbf r,t}$ is different in each layer of the network.

Note that when replacing~\eqref{eq:prox_reformulation} with~\eqref{eq:prox_param}, we have disentangled the proximity operator of $\tilde \psi$ and the derivative $\psi'$, in addition to introducing weights $\gamma^{(1)}_t$ and $\gamma^{(2)}_t$. As a result, the function $f_{\mathbf r}$ is no longer guaranteed to be a Bregman divergence, strictly speaking. Nevertheless, we can still interpret it as a measure of discrepancy between $\mathbf{y}$ and $\mathbf{r}$.

\section{Experiments}
\label{sec:exp}

In this section, we assess the potential of UADMM for PR of speech signals.
Our code is implemented using the PyTorch framework~\cite{pytorch} and will be made available online for reproducibility, as soon as the paper is accepted.
\subsection{Protocol}

\subsubsection{Data}
We build a set of speech signals from the TIMIT dataset~\cite{Garofolo1993}. The dataset is split into training, validation, and test subsets containing $1000$, $10$, and $50$ utterances, respectively. The signals are mono, sampled at $22,500$ Hz and cropped to 2 seconds. The STFT is computed with a 1024 samples-long (46 ms) self-dual sine window~\cite{vial2021phase} and 50$\%$ overlap. STFT magnitudes ($d=1$) are considered as nonnegative observations $\mathbf{r}$.

\subsubsection{Training}

\label{sec:training}
The network is trained with the ADAM algorithm~\cite{Kingma2015} using a learning rate of $10^{-4}$. We use a structure with $T=15$ layers and $C=3$, as these values have shown to be a good trade-off between performance and number of parameters in preliminary experiments. Batches of $10$ signals with a maximum of $200$ epochs are used for training. Training is stopped when the loss function on the validation subset starts increasing.
Given that we consider speech signals, we train the network by minimizing the negative short-term objective intelligibility measure (STOI)~\cite{STOI} between the estimated and ground truth signals. Indeed, this strategy was shown to be efficient for speech enhancement applications~\cite{Zhao2018,Naithani2018}. The negative STOI metric used for training the network is implemented in PyTorch via the \texttt{pytorch-stoi} library~\cite{pytorch-stoi}.

\subsubsection{Methods} As baselines, we consider GLA~\cite{GLA} (run for $1500$ iterations), and ADMM using a quadratic loss and ${\rho=10^{-3}}$, since this setup has exhibited good performance in our previous study under similar conditions~\cite{vial2021phase}. ADMM is run for a variable number of iterations, with a maximum of $1500$ (performance does not further improve beyond).
For fairness, the linear parts of UADMM use the same value for~$\rho$, and it is initialized such that $\mathsf F_t$ replicates the quadratic proximity operator (\textit{cf}. Section~\ref{sec:method_parametrization}).
All methods use the same initial signal estimate $\mathbf{x}_0$ computed using the ground truth magnitudes $\mathbf{r}$ and a random uniform phase, and $\bm \lambda_0=0$.

\subsubsection{Evaluation}
Reconstruction performance is assessed with the STOI metric (which ranges between $0$ and $1$, higher is better) computed on the test set with the \texttt{pystoi} library~\cite{pystoi}.

\subsection{Results}

\begin{figure}[t]
    \centering
    \includegraphics[width=0.4\linewidth]{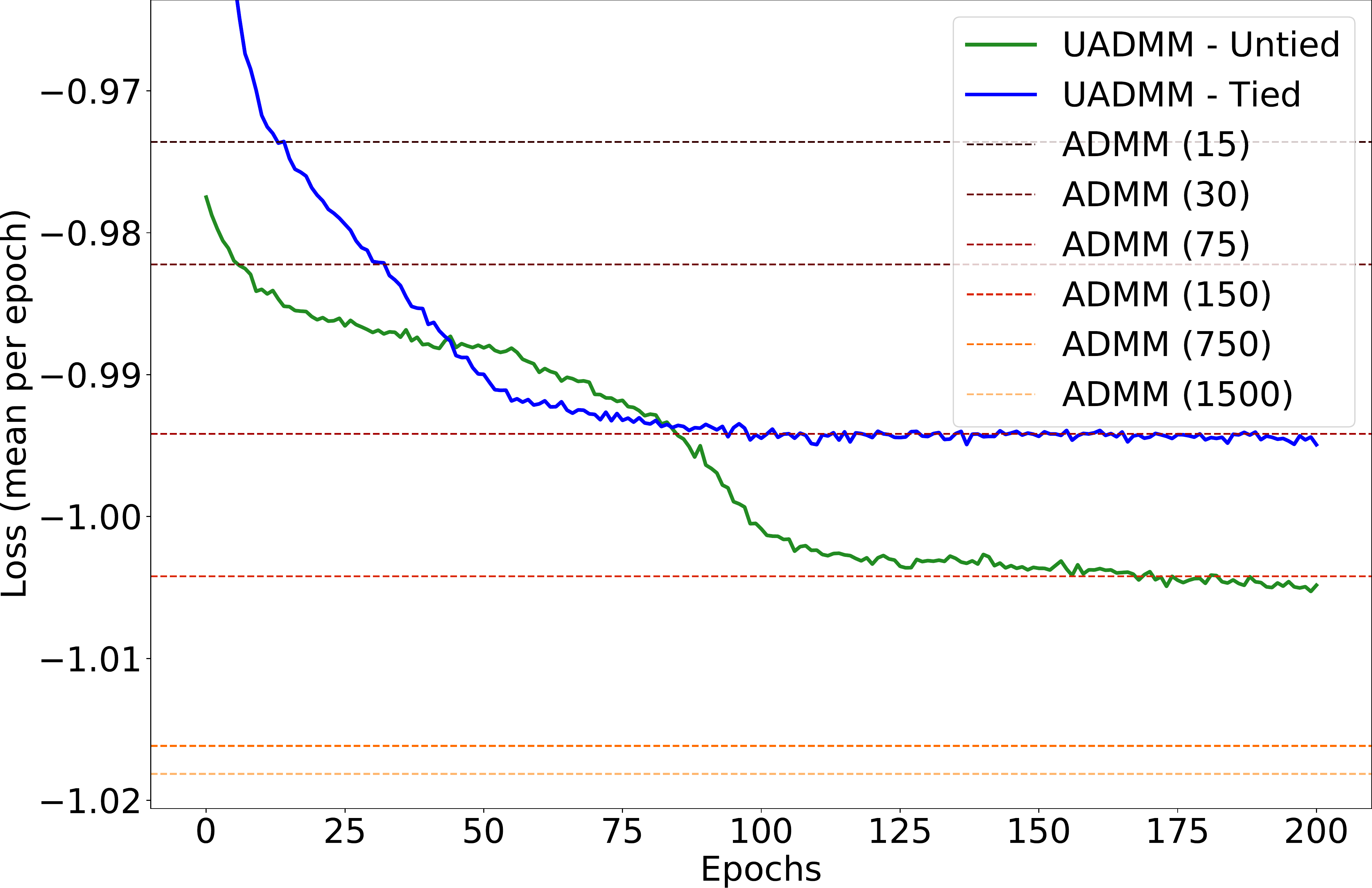}
    \caption{Training loss (negative STOI) over epochs. Note that \texttt{pytorch-stoi} implementation does not exactly replicate the original metric and consequently yields values lower than $-1$.}
    \label{fig:loss}
\end{figure}

\begin{figure}[t]
    \centering
    \includegraphics[scale=0.27]{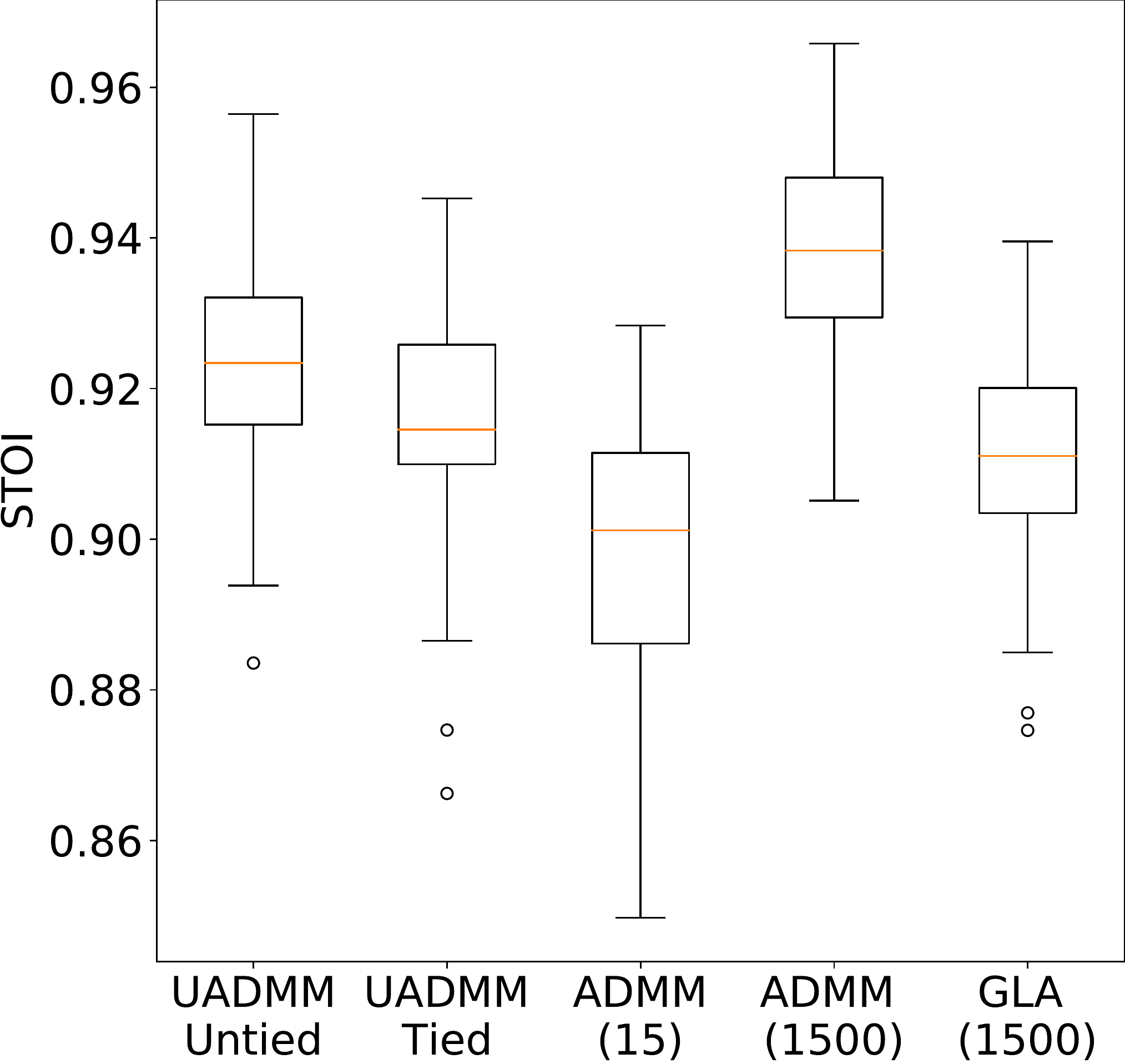}
    \caption{Performance on the test set. Each box-plot is made up of a central line indicating the median, box edges indicating the $1^{\text{st}}$ and $3^{\text{rd}}$ quartiles, whiskers indicating the extremal values, and circles representing the outliers.}
    \label{fig:eval}
\end{figure}

First, we display the training loss over epochs in Fig.~\ref{fig:loss}.
Both UADMM variants outperform the baseline ADMM with $15$ iterations on the training set.
UADMM-untied reaches a lower loss value than its tied counterpart, which was expected since this variant contains more learnable parameters. They reach a performance comparable to that of ADMM using $150$ and $75$ iterations, respectively.
The results on the test set presented in Fig.~\ref{fig:eval} confirm that the proposed UADMM approach significantly outperforms the classical ADMM using the same number of iterations, as well as the GLA baseline. A fully-converged ADMM algorithm (using $1500$ iterations) exhibits a higher STOI than our $15$ layers-based approach. Nonetheless, a more fair comparison would involve that both approaches use the same total number of iterations/layers. 

\begin{figure}[t]
    \centering
    \includegraphics[scale=0.25]{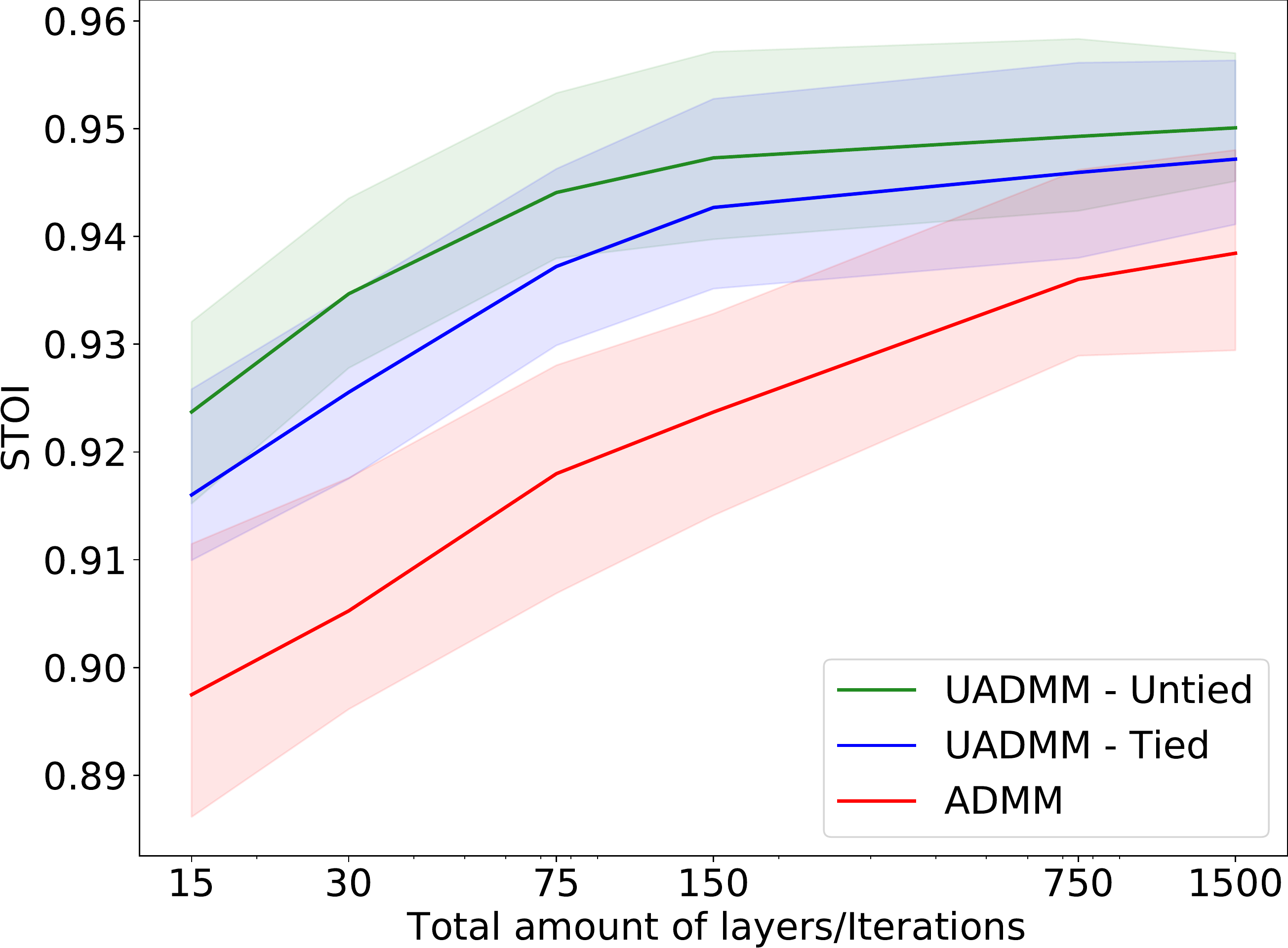}
    \caption{Evaluation with STOI over test dataset with iterated model. The solid lines denote the mean STOI and the the light colored areas the values between the first and the third quartile.}
    \label{fig:multi}
\end{figure}

\begin{figure}[t]
\centering
\includegraphics[scale=0.25]{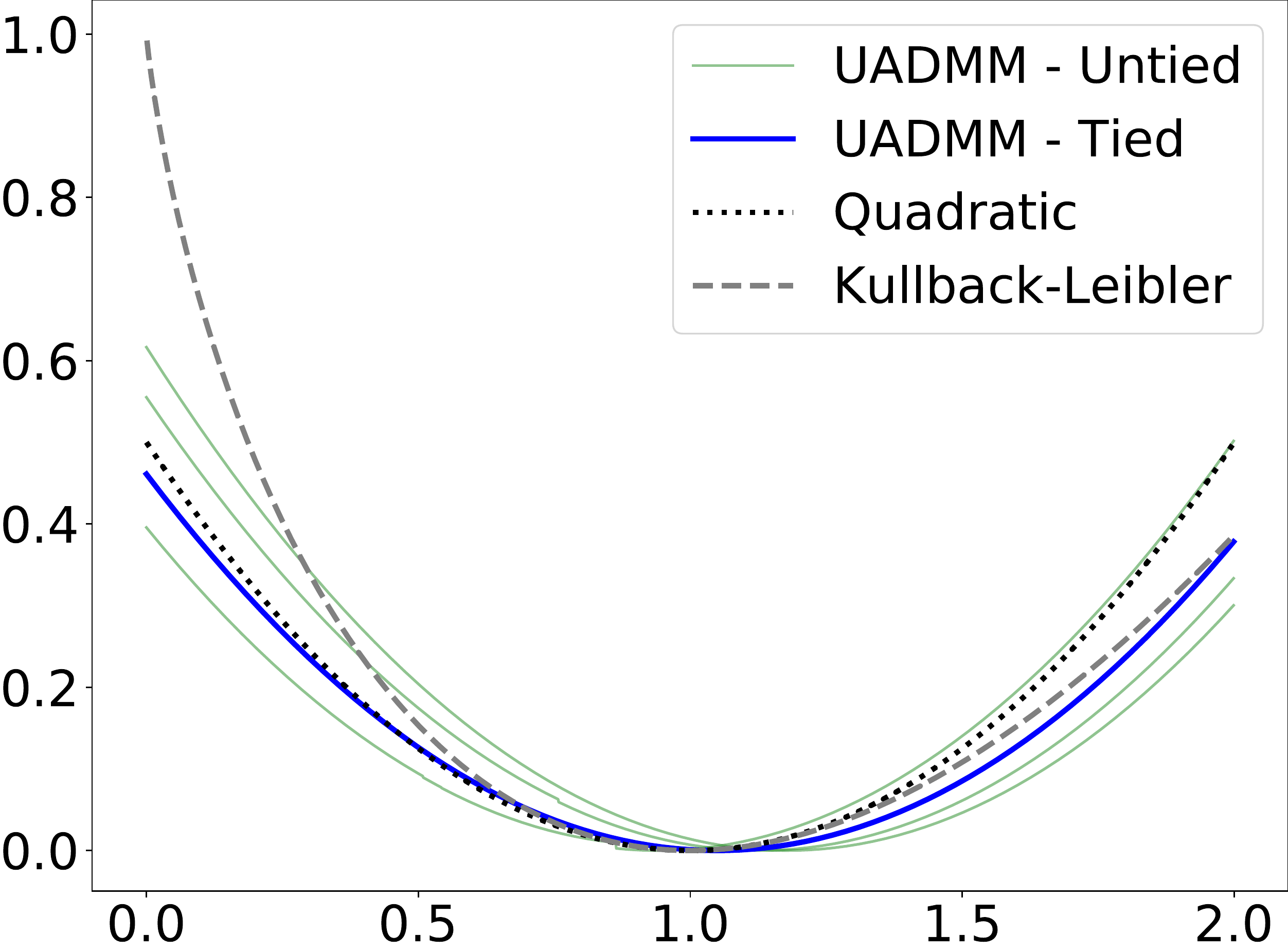}
\caption{Learned metrics $f_{r,t}(y)$  with $r = 1$. The quadratic loss and Kullback-Leibler divergence $\mathcal D_{KL}(y\,|\,r)$ are also displayed for the sake of comparison. In the ``tied" case, $f_r$ in analogous to $\mathcal D_\psi(\cdot\,|\, r)$ involved in the PR optimization problem. For clarity, only 3 of the 15 trained layers $f_{r,t}$ are displayed for the ``untied" case.}
\label{fig:divs}
\end{figure}

 To that end, we consider an \emph{ad hoc} variant of our method, where we apply the whole UADMM network several times: this allows to increase the total number of ``iterations" without the need for training a larger network from scratch.
The results presented in Fig.~\ref{fig:multi} show that this method consistently and significantly outperforms ADMM for any number of iterations. In particular, the performance of the fully-converged ADMM (after $1500$ iterations) is reached at only $30$ ``iterations'' for UADMM-untied (i.e., twice the number of trained layers), which exhibits the computational advantage of the proposed approach. Note that UADMM-tied with $T$ layers is equivalent to applying $T$ iterations of a standard ADMM algorithm using a learned metric $f_{\mathbf r}$.
We display these learned metrics ($f_{\mathbf r}$ in the tied case and $f_{\mathbf r,t}$ in the untied case) in Fig.~\ref{fig:divs}. These resemble beta-divergences with $\beta \in [1.5, 2.5]$. This is consistent with previous results from the literature~\cite{Fitzgerald2008}, where this range of values has shown good performance for audio spectral decomposition.

\section{Conclusion}
\label{sec:conc}

In this paper, we have addressed the problem of metric learning for phase retrieval by unfolding the recently proposed ADMM algorithm \cite{vial2021phase} into a neural network. We proposed to replace the proximity operator involved in this algorithm with learnable activation functions, since this operator conveys the information about the discrepancy measure used in formulating the PR problem. Experiments conducted on speech signals show that this approach outperforms the ADMM algorithm while keeping a light and interpretable structure.

\newpage
\bibliographystyle{IEEEbib}
\bibliography{references}

\newpage 

\renewcommand\thesubsection{\Alph{subsection}}

\newcommand{\labelsubseccounter}[1]{
    \renewcommand\thesubsection{\Alph{subsection}}
    \addtocounter{subsection}{-1}
    \refstepcounter{subsection}
    \label{#1}
}

\section*{Supplementary:  Characterization of $\mathsf F(\mathbf y, \mathbf r)$ as a proximity operator.}

In this supplementary, we address the problem of identifying a function $f_{\mathbf r}$ $: \mathbb R^K \rightarrow \mathbb R \cup \{+\infty\}$ such that ${\mathsf F(\mathbf y, \mathbf r) = \prox_{f_{\mathbf r}}(\mathbf y)}$, with $\mathsf F$ defined in~\eqref{eq:prox_param} of the main paper. Note that we ignore here the layer index $t$ for simplicity.

First, we present in Section~\ref{sec:apdx_prox} sufficient conditions for the existence of such a function in a general setting. Then, in Section~\ref{sec:apdx_APL} we consider more particularly the case of APL units. Finally, in Section~\ref{sec:apdx_F} we recover the function $f_{\mathbf r}$ corresponding to our sublayer $\mathsf F$.

\subsection{Proximity operators}
\labelsubseccounter{sec:apdx_prox}

First, let us recall that the proximity operator of a real-valued convex function $\sigma: \mathbb R^K \rightarrow \mathbb R \cup \{+\infty\}$ is defined as the mapping of a vector $\mathbf z$ to the solution of the minimization problem:
\begin{equation}
    \prox_{\rho^{-1}\sigma}(\mathbf z) := \argmin{\mathbf p \in \mathbb R^K} \sigma(\mathbf p) + \frac{\rho}{2}\|\mathbf p - \mathbf z\|^2.
\end{equation}
This definition can be extended to nonconvex functions, resulting in a possibly set-valued operator. Following \cite{moreau1965proximite, combettes2008proximal, gribonval2020characterization}, a function $g : \mathcal Z \subset \mathbb R^K \rightarrow \mathbb R^K$ can be characterized as the proximity operator of a convex function when it is strictly increasing over each coordinate. More precisely, the authors in~\cite{gribonval2020characterization} show that if there exists a convex, lower semi-continuous function $\tilde g: \mathbb R^K \rightarrow \mathbb R$ such that $g(\mathbf z)$ is a subgradient of $\tilde g(\mathbf z)$ for all~$\mathbf z \in \mathcal Z$, i.e.,
\begin{equation}
  \label{eq:gbar}
    g(\mathbf z)\in\{\mathbf v \in \mathbb R^K \,|\, \forall \mathbf p \in \mathbb R^K,\, \langle\mathbf p - \mathbf z,\, \mathbf v\rangle + \tilde g(\mathbf z) \leq \tilde g(\mathbf p) \},
\end{equation}
then there exists $\sigma$ such that
$g(\mathbf z) \in \prox_{\sigma}(\mathbf z)$
and the following relation stands:
\begin{equation}
\label{eq:proxrel}
    \forall \mathbf z \in \mathcal Z,\, \sigma(g(\mathbf z))=\langle \mathbf z,\, g(\mathbf z)\rangle - \frac 1 2 \|g(\mathbf z)\|^2 - \tilde g(\mathbf z).
\end{equation}
Finally, when $g$ is invertible and with $\mathbf y = g(\mathbf z)$, we have:
\begin{equation}
\label{eq:proxrelbis}
    \sigma(\mathbf y) = \langle g^{-1}(\mathbf y) ,\, \mathbf y \rangle - \frac 1 2 \|\mathbf y \|^2 - \tilde g(g^{-1}(\mathbf y)).
\end{equation}

\subsection{Characterization with APL}
\labelsubseccounter{sec:apdx_APL}

We now address the case of (strictly increasing) APL functions as defined in~\eqref{eq:apl}, with negative weights (\textit{cf}.~\ref{sec:method_interpretation}) and at least one nonnegative bias $b_c$. Let us consider the following convex, lower semi-continuous function $\overline{\APL}$ such that $\forall z \in \mathbb R$:
\begin{equation}
  \overline{\APL} (z)= \frac{z^2}{2}\chi_{[0;+\infty]}(z) \\+ \sum_{c=1}^C w_c \left( \frac{-z^2}{2}+b_c z \right) \chi_{]-\infty; b_c]}(z),
\end{equation}
where $\chi _\Pi$ is the indicator function of set $\Pi$. Since for any $z\in\mathbb{R}$, $\APL(z)$ is a subgradient of $\overline{\APL}(z)$, and denoting $\tildeAPL(\mathbf z) = \sum_{k=1}^K \overline{\APL} (z_k)$, it is straightforward to show that the relation~\eqref{eq:gbar} stands for ${g(\mathbf z)=\APL(\mathbf z)}$ and ${\tilde g(\mathbf z)= \tildeAPL(\mathbf z)}$.

Besides, since $\APL$ is invertible we can use the relation~\eqref{eq:proxrelbis} to identify~$\sigma$:
\begin{equation}
  \label{eq:sigma}
   \sigma(\mathbf y) = \langle \APL^{-1}(\mathbf y),\, \mathbf y \rangle - \frac 1 2 \|\mathbf y \|^2 - \tildeAPL(\APL^{-1}(\mathbf y)),
\end{equation}
with:
\begin{equation}
  \APL^{-1}(\mathbf y) = \frac{\mathbf y - \sum_{c=1}^C w_c b_c\chi_{]-\infty,\, \APL(b_c)]}(\mathbf y)}{\chi_{[\APL(0),\,+\infty[}(\mathbf y)- \sum_{c=1}^C w_c \chi_{]-\infty,\, \APL(b_c)]}( \mathbf y)}.
\end{equation}

\subsection{Characterization with $\mathsf F$}
\labelsubseccounter{sec:apdx_F}

Finally, let us retrieve $f_{\mathbf r}: \mathbb R^K \rightarrow \mathbb R \cup \{+\infty\}$  such that ${\mathsf F(\mathbf y, \mathbf r) = \prox_{f_{\mathbf r}}(\mathbf y)}$. Drawing on the previous section and using the definition of $\mathsf F$ from~\eqref{eq:prox_param}, we have:
\begin{equation}
    \mathsf F(\mathbf y, \, \mathbf r) = \text{prox}_{\sigma} \left(\gamma^{(1)}\mathbf y + \gamma^{(2)}\frac{\mathbf r^{\beta-1}}{\beta-1}\right).
    \label{eq:prox_sigma}
\end{equation}
To fully identify $f_{\mathbf r}$, we first need to reformulate~\eqref{eq:prox_sigma} so that the argument of the right hand side term simply becomes~$\mathbf{y}$. To that end, we leverage a property from~\cite{combettes2005signal}, which consists in first rewriting~\eqref{eq:prox_sigma} as follows:
\begin{equation}
    \mathsf F(\mathbf y, \, \mathbf r) = \text{prox}_{\sigma} \left(\frac{\mathbf y - \mathbf q}{2\alpha+1}\right),
    \label{eq:prox_refor}
\end{equation}
with $\alpha = \displaystyle \frac{1-\gamma^{(1)}}{2\gamma^{(1)}}$ and $\mathbf q = - \displaystyle \frac{\gamma^{(2)}}{\gamma^{(1)}} \displaystyle \frac{\mathbf r ^{\beta-1}}{\beta -1}$. The property from~\cite{combettes2005signal} then states that:
\begin{equation}
  \label{eq:proxprop}
  \text{prox}_{\varphi + \alpha\|\cdot\|^2 + \langle \mathbf q,\,\cdot \rangle} (\mathbf y)= \text{prox}_{\varphi/(2\alpha+1)}\left(\frac{\mathbf y - \mathbf q}{2\alpha+1}\right).
\end{equation}
Let $\varphi =  (2 \alpha +1) \sigma$. Combining~\eqref{eq:prox_refor} and~\eqref{eq:proxprop} yields:
\begin{equation}
    \mathsf F(\mathbf y, \, \mathbf r) = \text{prox}_{(2 \alpha +1) \sigma + \alpha\|\cdot\|^2 + \langle \mathbf q,\,\cdot \rangle} (\mathbf y).
    \label{eq:prox_f_sigma}
\end{equation}
As a result, from~\eqref{eq:prox_f_sigma} we can identify $f_{\mathbf r}$ such that its proximity operator is $\mathsf F$. If we further exploit the definition of $\sigma$ from~\eqref{eq:sigma}, we finally have:
\begin{equation}
  f_{\mathbf r}(\mathbf y) = \frac{1}{\gamma^{(1)}}\left \langle \APL^{-1}(\mathbf y) - \gamma^{(2)}\frac{\mathbf r ^{\beta-1}}{\beta -1} , \,\mathbf y \right \rangle \\ - \frac 1 2 \|\mathbf y\|^2 - \frac{1}{\gamma^{(1)}}\tildeAPL(\APL^{-1}(\mathbf y)).
  \label{eq:inv_prox}
\end{equation}
Therefore, using~\eqref{eq:inv_prox} one can recover the function associated with the learned proximity operator, and consequently identify the metric involved in the formulation of the PR problem.

\end{document}